\documentclass[prb,twocolumn,showpacs,amsmath,amssymb]{revtex4}

\usepackage{graphics}
\usepackage{dcolumn}
\usepackage{bm}

\def\zins  	{$\hbox{Zn}_{\hbox{In}}$ }
\def\zina  	{$\hbox{Zn}_{\hbox{In}}^{-1}$}

\def\vp  	{$\hbox{V}_{\hbox{P}}$}
\def\vps        {$\hbox{V}_{\hbox{P}}$ } 
\def\vpas       {$\hbox{V}_{\hbox{P}}^{+1}$ }
\def\vpas       {$\hbox{V}_{\hbox{P}}^{+1}$ }
\def\vpma       {$\hbox{V}_{\hbox{P}}^{-1}$}
\def\vpa        {$\hbox{V}_{\hbox{P}}^{+1}$}
\def\vpm        {\hbox{V}_{\hbox{P}}} 
\def\zvp        {$\left[\hbox{Zn}_{\hbox{In}}\hbox{-}\hbox{V}_{\hbox{P}}\right]$} 
\def\zvps       {$\left[\hbox{Zn}_{\hbox{In}}\hbox{-}\hbox{V}_{\hbox{P}}\right]$ } 
\def\zvpq       {$\left[\hbox{Zn}_{\hbox{In}}\hbox{-}\hbox{V}_{\hbox{P}}\right]^Q$} 
\def\zvpqs       {$\left[\hbox{Zn}_{\hbox{In}}\hbox{-}\hbox{V}_{\hbox{P}}\right]^Q$ }

\def\zvpz       {$\left[\hbox{Zn}_{\hbox{In}}\hbox{-}\hbox{V}_{\hbox{P}}\right]^0$} 
\def\zvpma      {$\left[\hbox{Zn}_{\hbox{In}}\hbox{-}\hbox{V}_{\hbox{P}}\right]^{-1}$}

\def\zvpmb      {$\left[\hbox{Zn}_{\hbox{In}}\hbox{-}\hbox{V}_{\hbox{P}}\right]^{-2}$ }
\def\zvpmc      {$\left[\hbox{Zn}_{\hbox{In}}\hbox{-}\hbox{V}_{\hbox{P}}\right]^{-3}$}
\def\zvpmd      {$\left[\hbox{Zn}_{\hbox{In}}\hbox{-}\hbox{V}_{\hbox{P}}\right]^{-4}$ }

\def\zvpmf      {$\left[\hbox{Zn}_{\hbox{In}}\hbox{-}\hbox{V}_{\hbox{P}}\right]^{-6}$ }
\def\zvppa      {$\left[\hbox{Zn}_{\hbox{In}}\hbox{-}\hbox{V}_{\hbox{P}}\right]^{+1}$}
\def\zvppb      {$\left[\hbox{Zn}_{\hbox{In}}\hbox{-}\hbox{V}_{\hbox{P}}\right]^{+2}$ }
\def\eform      {$E_{d}$} 
\def\eforms     {$E_{d}$ }

\begin{document}

\title{The Structure of the \zvps Defect Complex in Zn Doped InP.}

\author{C.W.M. Castleton}
	\affiliation{Theory of Condensed Matter, Department of Physics, Uppsala University,\\ 
	Box 530, 751 21 Uppsala, Sweden.}
 	\altaffiliation[Current address ]{Material Physics, Kungliga Techniska Hšgskolan, 229 KTH-Electrum, 16440 Kista,}
 	\email{Christopher.Castleton@fysik.uu.se}
\author{S.Mirbt} \affiliation{Theory of Condensed Matter, Department 
of Physics, Uppsala University,\\
	Box 530, 751 21 Uppsala, Sweden.}

\date{\today}

\begin{abstract}
We study the structure, the formation and binding energies and the 
transfer levels of the zinc-phosphorus vacancy complex \zvps in Zn 
doped p-type InP, as a function of the charge, using plane wave ab 
initio DFT-LDA calculations in a 64 atom supercell.  We find a binding 
energy of 0.39 eV for the complex, which is neutral in p-type 
material, the 0/-1 transfer level lying 0.50 eV above the valence band 
edge, all in agreement with recent positron annihilation experiments.  
This indicates that, whilst the formation of phosphorus vacancies 
(\vpa) may be involved in carrier compensation in heavily Zn doped 
material, the formation of Zn-vacancy complexes is not.

Regarding the structure: for charge states $Q=+6\rightarrow -4$ the Zn 
atom is in an $sp^2$ bonded DX position and electrons 
added/removed go to/come from the remaining dangling bonds on the 
triangle of In atoms.  This reduces the effective vacancy 
volume monatonically as electrons are added to the complex, also in 
agreement with experiment.  The reduction occurs through a combination 
of increased In-In bonding and increased Zn-In electrostatic 
attraction.  In addition, for certain charge states we find complex 
Jahn-Teller behaviour in which up to three different structures, (with 
the In triangle dimerised, antidimerised or symmetric) are stable and 
are close to degenerate.  We are able to predict and successfully 
explain the structural behaviour of this complex using a simple tight 
binding model.

\end{abstract}

\maketitle

\section{\label{Intro}Introduction}

Zn doped p-type InP is one of the most common materials in use within 
optoelectronics. The Zn sits substitutionally
within the In sublattice ($\hbox{Zn}_{\hbox{In}}$) 
where it has a shallow acceptor level. A well known 
limitation in the use of 
Zn as a p-dopant in InP is the saturation of the hole concentration 
in the mid 10$^{18}$ cm$^{-3}$ range.  Above this further increases in 
Zn concentration do not translate into increases in hole 
concentration.  There are several suspected causes\cite{Mahony} for 
this, in particular: a) increases in the concentration of interstitial 
zinc ($\hbox{Zn}_{\hbox{i}}$), b) phase separation and c) the formation 
of other compensating defects, especially phosphorus 
vacancies\cite{Mahony} (\vp) and complexes of zinc with \vp.  In this 
paper we will examine the properties of Zn-\vps complexes, using ab 
initio methods.

The formation of one such Zn-\vps complex has recently been 
studied\cite{Slotte,Saarinen2} by Slotte et al.  in Zn doped 
Czochralski grown samples which have been post-growth annealed in the 
temperature range 300-600 K.  Using positron annihilation they find 
that the complex has at least two forms.  The most stable at room 
temperature\cite{Slotte} has an effective vacancy volume 
larger\cite{Alatalo} than that of a free \vps.  A binding energy on 
the order of $0.1\rightarrow 0.4$ eV is anticipated\cite{Saarinen2} 
(prior to detailed measurements).  When the measurement temperature is 
raised they find that the complex undergoes a transition to a form 
with a smaller vacancy volume, similar to that of the free vacancy.  
It has been suggesed that the more stable form is neutral\cite{Slotte} 
and has a DX structure\cite{Saarinen2} which becomes non-DX when 
excited to the -1 charge state.  They find\cite{Slotte} the 0/-1 
transfer level 0.2$\pm$0.1 eV above the valence band edge.

A number of theoretical studies of the free phosphorus vacancy have 
been performed\cite{EarlyVP,Nieminen1,Nieminen2} using Density 
Functional Theory (DFT).  The free vacancy has been 
shown\cite{Nieminen1,Nieminen2} to be a strong Jahn-Teller defect, 
indeed, to be a ``negative U'' centre in which the neutral charge 
state $\vpm^{0}$ is thermodynamically unstable relative to the +1 and 
-1 states.  Hence a direct change $\vpm^{-1} \rightarrow \vpm^{+1}$ 
occurs as a function of increasing electron chemical potential (Fermi 
level).  This is due to strong lattice relaxation in the -1 charge 
state.  The vacancy is surrounded by a tetrahedron of four In atoms, 
each with a (partially filled) inwardly pointing dangling bond.  In 
the -1 charge state a pair of dimers is formed: two opposing sides of 
the tetrahedron are reduced in length relative to the remaining four.  
The energy gained by this overcomes the Coulomb repulsion between the 
additional electrons, allowing two electrons to be transfered onto the 
vacancy simultaneously.
 
In this paper we present what is, to our knowledge, the 
first ab initio\footnote{Alatalo et al.  have\cite{Alatalo,Slotte} used 
DFT-LDA and DFT-GGA calculations to estimate the positron annihilation 
at Zn-\vps complexes, but they included only empiracle breathing mode 
relaxations (to fit the positron lifetimes) and presented no results 
for the structure, transfer levels or energetics of the complex.} ab 
initio theoretical study of the \zvps complex, which we propose to be 
that observed in the positron annihilation experiments.  We will first 
describe the method in section \ref{Method}.  We will then report 
detailed results for the geometry and vacancy volume of the complex, 
both for the ground state structures and for various metastable 
structures in section \ref{Q=0 Str}, followed by a description of how 
these vary with changes in the charge state of the complex in section 
\ref{All Q str}.  In section \ref{model} we will also propose a simple 
tight binding model which is able to predict almost all of the 
structural properties which we find.  Finally, in section 
\ref{conclusions} we will conclude, relating our results to those from 
position annihilation and the issue of Zn compensation.

\section{\label{Method}Calculational Details.}

We use planewave {\it ab initio} DFT\cite{DFT} within the Local Density 
Approximation (LDA) together with ultrasoft 
pseudopotentials\cite{Vanderbilt,Kresse}.  Calculations are performed 
using the VASP code\cite{VASP}.  Before doing calculations involving 
defects we first optimise the lattice parameter subject to LDA.  We 
find a value of 5.827 \AA, compared to 5.869 \AA~experimentally, 
giving us a band gap of 0.667 eV, compared to 1.344 eV in experiment.

\begin{table}
\caption{\label{kconv}Comparison of the k-point convergence of the 
electronic and structural contributions to the formation energy, 
\eform.  Convergence of the electronic contributions is given by the 
change, $\Delta^{UR}(\hbox{N-M})$, in the unrelaxed 
formation energy when increasing the k-point grid from NxNxN to and 
MxMxM.  Convergence of the structural contributions is given by 
$\Delta^{R}(\hbox{N-M})$, the equivalent change in 
relaxation energy $\epsilon_R(\hbox{N})$.  See main text for 
full definitions.}
\begin{ruledtabular}
\begin{tabular}{lrrr} 
Defect                    &$\Delta^{UR}(\hbox{2-4})$ &$\Delta^{UR}(\hbox{6-4})$ &$\Delta^R(\hbox{2-4})$ \\\hline
$\left.\hbox{V}_{\hbox{In}}\right.^{-4}$    		&  0.0082 	&		&-0.0190	\\	
$\left.\hbox{V}_{\hbox{In}}\right.^{0}$    		& -0.0483	&0.0093		& 0.0368	\\
$\left.\hbox{In}_{\hbox{i}}\right.^{0}$			&  0.0252	&0.0052		& 		\\
$\left.\hbox{In}_{\hbox{i}}\right.^{+1}$		& -0.0539	&		&-0.0037	\\ 
$\left.\hbox{In}_{\hbox{i}}\right.^{+3}$		&  0.0206	&0.0058		&  		\\
$\left.\hbox{In}_{\hbox{i}}\right.^{+4}$		& -0.1038	&		& 0.0061	\\
$\left.\hbox{P}_{\hbox{In}}\right.^{+1}$		& -0.0360	&		& 0.0246	\\
$\left.\hbox{P}_{\hbox{In}}\right.^{+2}$		&  0.0314	&		& 0.0259	\\
$\left.\hbox{V}_{\hbox{P}}\right.^{-2}$			& -0.0607 	&-0.0123	&  		\\
$\left.\hbox{V}_{\hbox{P}}\right.^{+2}$			&  0.0840	& 0.0012	&  		\\
$\left.\hbox{Zn}_{\hbox{In}}\right.^{+1}$		&  0.0375	&		& 0.0007	\\
$\left.\hbox{Zn}_{\hbox{In}}\right.^{0}$		&  0.0807	&-0.0011	&-0.0025	\\
$\left.\hbox{Zn}_{\hbox{In}}\right.^{-1}$		& -0.0004	& 0.0000	&-0.0005	\\
$\left.\hbox{Zn}_{\hbox{P}}\right.^{+2}$		& -0.0266	&		& 0.0095	\\
$\left.\hbox{Zn}_{\hbox{i}}\right.^{+2}$		&  0.0117	& 0.0001	& 0.0040	\\
$\left.\hbox{Si}_{\hbox{In}}\right.^{0}$		&  0.0091	&		& 0.0091	\\
$\left.\hbox{Si}_{\hbox{In}}\right.^{+1}$		&  0.0048	&		&-0.0038	\\
$\left.\hbox{Si}_{\hbox{In}}\right.^{+2}$		&  0.0891	&		& 0.0046	\\
$\left.\hbox{Si}_{\hbox{P}}\right.^{-2}$		&  0.0296	&		&-0.0006	\\
$\left.\hbox{Si}_{\hbox{P}}\right.^{-1}$		&  0.0031	&		& 0.0008	\\
$\left.\hbox{Si}_{\hbox{P}}\right.^{ 0}$		&  0.0734	&		&-0.0027	\\
$\left.\hbox{Si}_{\hbox{P}}\right.^{+1}$		&  0.0279	&		& 0.0004	\\
$\left[\hbox{Zn}_{\hbox{i}}\hbox{-}\hbox{V}_{\hbox{P}}\right]^{+3}$&0.0272&		& 0.0151	\\
\zvpma							& -0.0750	&		& 0.0046	\\
\zvpz							&  0.0057	&		&-0.0029	\\
\zvppa							& -0.0598	&		& 0.0184	\\
\end{tabular}
\end{ruledtabular}
\end{table}

For defect calculations we use
a 64 atom simple cubic supercell and allow all atoms not located on 
the surface of the cell to relax.  This restriction is included to 
truncate the (spurious) elastic interactions between adjacent 
supercells.  No restrictions are placed upon the symmetry of the 
relaxations.  When calculating total energies of supercells containing 
charged defects we also include a uniform neutralising background 
charge.  The key quantity is the formation energy of the defect, 
\eform, defined\cite{eformation} as
\begin{equation}
E_d = E^T(\hbox{defect}^Q) - E^T(\hbox{bulk}) + \sum_i \mu_in_i 
+ Q\left( e_v + e_f \right)
\end{equation}
where $E^T(\hbox{defect}^Q)$ and $E^T(\hbox{bulk})$ are the 
total energy of the InP supercell with and without the charge $Q$ 
defect.  Both are calculated with the same values of planewave cutoff, 
k-point grid, etc, in order to make use of cancellation of errors.  
The defect is formed by adding(removing) $n_i$ atoms, each with 
chemical potential $\mu_i$, and by adding ($-Q$) electrons, whose 
chemical potential $e_f$ is measured from the valence band edge $e_v$.  
We then adjust the various calculational parameters to converge 
\eform.  We find that it is sufficient to use pseudopotentials in 
which Zn d-electrons are treated as valence electrons but In 
d-electrons are left in the core.  We find that a planewave cutoff 
energy of 200 eV is enough to converge \eforms to $O( 0.01 \hbox{ 
eV})$.  Real space projection operators are used, with the 
cutoff again chosen to give errors below $O( 0.01 \hbox{ eV})$.

The k-point convergence is more interesting, since we find different 
convergence behaviour for the electronic and structural contributions to 
the formation energy $E_d(\hbox{K})$, where we use a KxKxK 
Monkhorst-Pack\cite{MP} k-point grid.  The error in the electronic 
contribution to $E_d(\hbox{K})$ is very close to the error in the 
unrelaxed formation energy, $E_d^{UR}(\hbox{K})$, which we 
can estimate as
\begin{equation}
    \Delta^{UR}(\hbox{N-M}) = E_d^{UR}(\hbox{N}) - 
    E_d^{UR}(\hbox{M})
\end{equation}
(We shall use \{N,M\} = \{2,4\} and \{4,6\} to estimate the errors in 
$E_d^{UR}(\hbox{2})$ and $E_d^{UR}(\hbox{4})$
respectively.) We cannot, of course, 
calculate the structural contribution directly, so we first define the 
relaxation energy
\begin{equation}
\epsilon_R(\hbox{K}) = E_d^R(\hbox{K}) - E_d^{UR}(\hbox{K})
\end{equation}
where
$E_d^R(\hbox{K})$ is the relaxed formation energy.  We then 
estimate the convergence of $\epsilon_R(\hbox{K})$ using
\begin{equation}
    \Delta^{R}(\hbox{N-M}) = \epsilon_R(\hbox{N}) - \epsilon_R(\hbox{M})
\end{equation}
In table \ref{kconv} we list values of $\Delta^{UR}(\hbox{2-4})$, 
$\Delta^{UR}(\hbox{4-6})$ and 
$\Delta^{R}(\hbox{2-4})$ for a range of charge states 
of various different defects in InP.
We find that when atomic relaxations are ignored, a 2x2x2 k-point 
grid produces errors below $O( 0.01 \hbox{ eV})$, provided 
$Q=0$.  (Column $\Delta^{UR}(\hbox{2-4})$ of table \ref{kconv}.) When 
$Q\neq0$, however, we find that the error introduced by truncating at 
2x2x2 increases with $\mid$$Q$$\mid$.  Instead, a 4x4x4 grid is 
required: column $\Delta^{UR}(\hbox{4-6})$ indicates that this is 
sufficiently converged.  On the other hand, the relaxation energy 
$\epsilon_R(\hbox{K})$ converges faster, see column 
$\Delta^{R}(\hbox{2-4})$ in the table.  Indeed, $\epsilon_R(\hbox{K})$ 
is converged for a 2x2x2 Monkhorst-Pack grid, for any value of $Q$.  
In other words the structural contribution to \eforms appears to 
converge faster with k-point grid than the electronic contributions.  
This is largely due to a double cancellation of errors when we 
calculate $\epsilon_{R}$ from values of $E_{d}$ which already contain 
a cancellation of errors.  It means that, although we should use a 
4x4x4 grid to obtain an accurate value for $E_{d}^{R}$, we can save 
much calculation time by estimating $E_{d}^{R}(\hbox{4})$ as
\begin{eqnarray}
E_d^R(\hbox{4}) & \approx & E_d^{UR}(\hbox{4}) + \epsilon_{R}(\hbox{2}) \nonumber\\
      & = & E_d^{UR}(\hbox{4}) + E_d^R(\hbox{2}) - E_d^{UR}(\hbox{2})\qquad
\label{ktrick}
\end{eqnarray}
The resulting values of $E_d^R(\hbox{4})$ are converged to at 
least 0.02-0.04 eV, and usually to an order of magnitude or so better.  
Hence we do most relaxations with a 2x2x2 k-point grid, only checking 
key ones at 4x4x4.  Charge and Electron Localisation 
Function\cite{ELF} (ELF) plots are, on the other hand, calculated 
using a 4x4x4 k-point grid at the 2x2x2 related structures.

\begin{table}
\caption{\label{Levels}Transfer levels of \zvp, in the 64 atom 
cell, relative to the valence band edge.}
\begin{ruledtabular}
\begin{tabular}{cr}
$Level$	&Energy 	\\\hline
+6/+5 &-0.51 eV \\
+5/+4 &-0.51 eV \\
+4/+3 &-0.47 eV \\
+3/+2 &-0.44 eV \\
+2/+1 &-0.37 eV	\\
+1/0  &-0.32 eV	\\
0/-1  &0.50 eV	\\
-1/-2 &0.70 eV	\\
-2/-3 &0.95 eV	\\
-3/-4 &1.05 eV  \\
-4/-5 &1.19 eV  \\
-5/-6 &1.18 eV  \\
\end{tabular}
\end{ruledtabular}
\end{table}
\begin{table}
\caption{\label{Q str table}Geometry of the most stable form of \zvpqs 
as a function of $Q$.  Distances in \AA, volumes in \AA$^3$.  Most 
structures (Str.) found are symmetric (SY) (all In-In distances equal, 
all Zn-In distances equal.) Where the In-In (and likewise Zn-In) 
distances differ it is found that two remain equal (In-In(2)) and one 
is different (In-In(1)) by amount $\Delta(In)$.  
$(In-In(1))<(In-In(2))$ corresponds to the formation of a dimer (DM) 
and $(In-In(1))>(In-In(2))$ to an antidimer (AD).}
\begin{ruledtabular}
\begin{tabular}{rrrrrrrrr}
$Q$                &Zn-P   &\multicolumn{2}{c}{Zn-In}&\multicolumn{2}{c}{In-In}&Volume &Str.&$\Delta$(In)\\
                   &       &(1)      &(2)            &(1)      &(2)            &       &    &        \\\hline
+6                 &2.29   &\multicolumn{2}{c}{4.97} &\multicolumn{2}{c}{4.95} &14.36  &SY  &        \\
+5                 &2.29   &\multicolumn{2}{c}{4.97} &\multicolumn{2}{c}{4.96} &14.42  &SY  &        \\
+4                 &2.29   &\multicolumn{2}{c}{4.98} &\multicolumn{2}{c}{4.97} &14.49  &SY  &        \\
+3                 &2.29   &\multicolumn{2}{c}{4.97} &\multicolumn{2}{c}{4.99} &14.53  &SY  &        \\
+2                 &2.29   &\multicolumn{2}{c}{4.99} &\multicolumn{2}{c}{4.97} &14.57  &SY  &        \\
+1                 &2.30   &\multicolumn{2}{c}{4.69} &\multicolumn{2}{c}{4.36} &10.86  &SY  &        \\
+0                 &2.30   &\multicolumn{2}{c}{4.41} &\multicolumn{2}{c}{3.69} &7.58   &SY  &        \\
-1                 &2.30   &4.23     &4.31           &3.64     &3.37           &6.52   &AD  &7.9\%   \\
-2                 &2.31   &4.13     &4.20           &3.52     &3.19           &5.81   &AD  &10.3\%  \\
-3                 &2.32   &3.95     &4.10           &2.99     &3.28           &5.23   &DM  &8.9\%   \\
-4                 &2.32   &\multicolumn{2}{c}{3.96} &\multicolumn{2}{c}{3.08} &4.85   &SY  &        \\
-5                 &2.43   &\multicolumn{2}{c}{3.21} &\multicolumn{2}{c}{3.04} &3.59   &SY  &        \\
-6                 &2.51   &\multicolumn{2}{c}{2.99} &\multicolumn{2}{c}{2.95} &3.09   &SY  &        \\
\end{tabular}
\end{ruledtabular}
\end{table}

\begin{table}
\caption{\label{Metastr table}Geometry of structural excitations 
of \zvpqs as a function of $Q$.  Energies are given relative to the 
energy of the most stable structures, as listed in table \ref{Q str 
table}.  See table \ref{Q str table} also for meaning of other 
collumns and notation.}
\begin{ruledtabular}
\begin{tabular}{rlrrrrrrrr}
$Q$                &Energy&Zn-P   &\multicolumn{2}{c}{Zn-In}&\multicolumn{2}{c}{In-In}&Volume &Str.&$\Delta(In)$ \\
                   &[eV]  &       &(1)      &(2)            &(1)      &(2)            &       &    &        \\\hline
+4                 &1.00  &2.30   &\multicolumn{2}{c}{4.44} &\multicolumn{2}{c}{3.74} &7.85   &SY  &        \\
+3                 &0.97  &2.30   &\multicolumn{2}{c}{4.44} &\multicolumn{2}{c}{3.76} &7.91   &SY  &        \\
+2                 &0.93  &2.30   &\multicolumn{2}{c}{4.44} &\multicolumn{2}{c}{3.72} &7.76   &SY  &        \\
+1                 &0.12  &2.30   &\multicolumn{2}{c}{4.42} &\multicolumn{2}{c}{3.71} &7.67   &SY  &        \\
-1		   &0.011 &2.30   &4.29     &4.27           &3.41     &3.50           &6.57   &DM  &2.7\%   \\
-1                 &0.012 &2.30   &\multicolumn{2}{c}{4.28} &\multicolumn{2}{c}{3.47} &6.58   &SY  &        \\
-2                 &0.03  &2.31   &4.18     &4.17           &3.22     &3.35           &5.85   &DM  &3.7\%   \\
-3                 &0.02  &2.31   &4.05     &4.10           &3.41     &3.06           &5.26   &AD  &11.6\%  \\		   
-3                 &0.05  &2.31   &\multicolumn{2}{c}{4.08} &\multicolumn{2}{c}{3.18} &5.31   &SY  &        \\		   
-5                 &0.02  &2.34   &\multicolumn{2}{c}{3.67} &\multicolumn{2}{c}{3.05} &4.32   &SY  &        \\		   
-6                 &0.19  &2.34   &\multicolumn{2}{c}{3.63} &\multicolumn{2}{c}{3.05} &4.26   &SY  &        \\		   
\end{tabular}
\end{ruledtabular}
\end{table}

\section{\label{Q=0 Str}The neutral \zvps complex.}

\subsection{Formation and Binding Energies.}

In order to calculate absolute values for the formation and binding 
energies we need to know the chemical potentials $\mu_{\hbox{In}}$, 
$\mu_{\hbox{P}}$ and $\mu_{\hbox{InP}}$.  Fully converged calculations 
for In, P (black phosphorus) and InP (the latter in its basic 2 atom 
FCC cell) give the bulk values\footnote{Bulk values are calculated 
using the same pseudopotentials as we use for the defect 
calculations, but they are each fully converged in terms of both k-point 
grid and planewave cutoff.} -3.270 eV, -6.028 eV and -9.728 eV 
respectively.  For pure InP we have, at thermodynamic 
equilibrium,
$\mu_{\hbox{InP}}=\mu_{\hbox{In}}+\mu_{\hbox{P}}$.  In In rich 
material we then have $\mu_{\hbox{In}}$ = -3.270 eV leading to 
$\mu_{\hbox{P}}$ = -6.459 eV, whilst in P rich material 
$\mu_{\hbox{P}}$ = -6.028 eV and hence $\mu_{\hbox{In}}$ = -3.700 eV.  
We now use the values $\mu_{\hbox{In}}$ = -3.485 eV and 
$\mu_{\hbox{P}}$ = -6.243 eV, corresponding to exactly stoichiometric 
conditions.  Properly, doping with Zn should alter these $\mu$ values, 
but even a Zn concentration of 5.10$^{18}$ cm$^{-3}$ only corresponds 
to replacing 1 In in 2000 with Zn.  We thus ignore the effect, and use 
the fully converged bulk value -1.891 eV for $\mu_{\hbox{Zn}}$.

Since we are primarily interested in strongly p-type material, we 
assume that the electron chemical potential lies at the valence band 
edge, meaning that $e_f=0$.  Using fully relaxed calculations in the 
64 atom simple cubic cell, we find that the most stable charge states of 
the \zins and \vps are -1 and +1 respectively, whilst for \zvps it is 
0.  The formation energies are $E_d$(\zina) = 0.32 eV, $E_d$(\vpa) = 
1.96 eV and $E_d$(\zvpz) = 1.89 eV.  For \vps we find a free vacancy 
volume of 6.5 \AA$^3$, where we calculate only the volume of the 
tetrahedron formed by the four In surrounding the vacancy.  \vpas is 
invisible to positrons, so we will also report here the volume of 
\vpma, which is 4.6 \AA$^3$.  More detailed results about \vps and 
\zins will appear elsewhere.\cite{OurVP}

We can evaluate the binding energy, $E_{b}$, of the complex 
as
\begin{equation}
  E_b(\left[\hbox{Zn}_{\hbox{In}}\hbox{-}\hbox{V}_{\hbox{P}}\right]^0) = 
E_d(\hbox{Zn}_{\hbox{In}}^{-1}) + E_d(\hbox{V}_{\hbox{P}}^{+1}) -
E_d(\left[\hbox{Zn}_{\hbox{In}}\hbox{-}\hbox{V}_{\hbox{P}}\right]^0) 
\end{equation}
This gives us a value of 0.39 eV, which agrees with the 
$0.1\rightarrow0.4$ eV estimate from the positron annihilation 
experiments\cite{Saarinen2}.

\subsection{Structure.}

In figure \ref{fig Q=0_st} we show the relaxed structure of the \zvps 
complex in the $Q=0$ charge state.  The \zins is bonded to three P 
atoms only and has relaxed back into a DX-like position in the same 
plane as those P atoms.  The resulting Zn-P distances are 2.31 \AA, 
which is almost identical to the 2.34 \AA~value we find for the 
shorter Zn-P distances in a similar DFT-LDA calculation for 
Zn$_3$P$_2$ and various other Zn-P binary compounds.  \cite{ZnP} The three 
In atoms on the other side of the \vps have moved together slightly, 
giving three equal In-In distances of 3.69 \AA, compared to an ideal 
LDA bulk next nearest neighbour distance of 4.12 \AA.  The Zn-In 
distance is rather longer than the bulk value, however, at 4.41 \AA.  
This gives a volume for the vacancy of 7.58 \AA$^3$, where we 
calculate simply the volume of the tetradehron formed by the three In 
and the Zn.  In the remaining figures most of the atoms in the cell 
are omitted for clarity, and lines are drawn around the edges of the 
tetrahetrdon to guide the eye.  In the case of Zn-In lines they do 
{\it not} represent bonds.  This can be seen clearly by examining the 
ELF\cite{ELF} plots in figure \ref{fig Q=0_elf}.  There is no ELF 
density at all between the Zn and the In triangle, indicating that 
there is no bonding between them.  Instead, the Zn has three $sp^2$ 
hybrid orbitals bonded to the surrounding P, these bonds being filled.  
There is no density related to the Zn at all along the line 
perpendicular to the Zn-P plane (the (111) axis).  This indicates that 
the single Zn $2p^z$ orbital left over from the formation of the 
$sp^2$ hybrids is empty.

\section{\label{All Q str}The structure of \zvps as a function of 
charge state.}

\subsection{Bond lengths and vacancy volumes.}

In table \ref{Levels} we give the positions of a number of transfer 
levels of the complex.  The structures of all of the states we have 
examined (+6 $\rightarrow$ -6) are very similar, as listed in table 
\ref{Q str table}.  For all $Q>-5$ the Zn lies in the DX-like 
position.  The three In atoms remain unbonded to Zn, but both the 
In-In distances and the Zn-In distances shorten with increasing 
negative charge.  This reduces the vacancy volume for each additional 
electron.  This holds until $Q=-5$, when the Zn $2p^z$ orbital starts 
to be filled, leading at $Q=-6$ to a non-DX position for the Zn atom, 
with the Zn-In distance only 1.6\% longer than the In-In.  The 
transfer level into this state lies 1.18 eV above the valence 
band edge.  With the LDA band gap being only 0.67 eV this places the 
level about 0.51 eV above the conduction band edge.  This orbital also 
displays a negative $U$ effect: the $Q=-5$ structure - which lies half 
way between DX and non-DX forms - is thermodynamically unstable, so 
the transfer level -5/-6 lies just below the -4/-5 level.  Hence we 
anticipate instead a -4/-6 level about 0.5 eV above the conduction 
band edge.

In addition to this, for $Q=-1\rightarrow -3$, Jahn-Teller effects 
are apparent, in which one In-In distance is shortened (dimerised) or 
lengthened (antidimerised) relative to the other two.  These 
remaining In-In distances always remain equal to at least 
0.001\AA. 

In table \ref{Metastr table} we show the various 
metastable structures (structural excitations) which we find.  (An 
extensive search suggests that there are no other metastable structures 
besides these.\footnote{Certain additional structures - 
a dimerised structure for \zvppa, dimerised and antidimerised 
structures for \zvpmd and a second weaker dimer structure for \zvpmb - 
are found when the 2x2x2 Monkhorst-Pack k point grid is used.  
However, these turn out to be unstable when recalculated using a 4x4x4 
k point grid.} The 
fact that the Zn $2p^z$ level lies well inside the conduction band is 
further underlined by the metastable structures seen for $Q=-5$
and $Q=-6$.  For both of these we find a DX-like 
metastable structure very similar to the ground state structure of the 
$Q=-4$ state.  This indicates that one or two electrons (respectively) 
have been transfered from the complex into the conduction band, 
leaving the complex with an empty Zn $2p^z$ and additional charge 
elsewhere in the supercell.  The energy splitting between these DX and 
non-DX forms is therefore expected to change significantly (indeed, to 
switch over) in larger supercells where less defect level dispersion 
is seen.  A similar effect is seen in the $Q=+1\rightarrow+4$ charge 
states, where a metastable structure is found which closely resembles 
the $Q=0$ structure.  Since the hole levels corresponding to these 
charge states lie below the valence band edge, these structural 
excitations amount to exciting holes off the complex, leaving it in 
the $Q=0$ charge state, with additional charge elsewhere.  (An 
equivalent structural excitation for $Q=+5$ or +6 has not been found - 
presumably the excitation energy involved is too great for even 
metastability within this supercell.)

All of the other excited structures found in the calculations relate to 
variations in Jahn-Teller distortions and affect only the In-In and 
Zn-In bond lengths.  They involve no significant changes in vacancy 
volume, Zn position or electron occupancy of the 
complex (see next subsection).  The energies involved are very small indeed - on or below 
the order of room temperature thermal energies.  Room temperature 
experiments are therefore likely to see averages over these 
structures.  Three possible structures are found to be stable - one 
In-In bond dimerised (the other two equal), one antidimerised or all 
equal.  How the symmetric structure comes to be stable despite the 
Jahn-Teller effect will be returned to in section \ref{model}.  All 
three structures are found to be stable for $Q=-1$ and $Q=-3$, but for 
$Q=-2$ the symmetric structure is not seen.  For $Q=0$ and +1 we find 
only the symmetric structure.

The energy separations between these structures are on the limits of 
what we expect to resolve using only the 2x2x2 Monkhorst-Pack k-point 
grid together with equation \ref{ktrick}.  As a result we have 
re-calculated the energies and structures of all of these states 
using a 4x4x4 k-point Monkhorst-Pack grid, and, for some, even a 6x6x6 
grid.  We find that all of the structures reported in table \ref{Q str 
table} and \ref{Metastr table} remain stable, although the degree of 
dimerisation or antidimerisation changes in some cases.  The order of 
stability for the three structures changes for $Q=-1$ and $Q=-3$, 
however, and in all charge states they move closer to degeneracy.  
These changes are illustrated in table \ref{Excite ab initio 
table}.

\begin{table}
\vglue -0.3truecm
\caption{\label{Excite ab initio table}Ab initio results for the 
Lowest Lying structures (marked L.L.) of \zvpqs for $Q=+2\rightarrow-4$, 
together with structural excitation energies in electron volts.  DM = 
dimerised, SY = symmetric, AD = antidimerised.  ''-'' indicates that 
the structure is not stable.}
\begin{ruledtabular}
\begin{tabular}{rccc}
Q       &DM             &SY             &AD                 \\\hline
+2      &-              &L.L.           &-                  \\
+1      &-              &L.L.           &-                  \\
0	&-		&L.L.           &-	            \\
-1      &L.L.           &0.0004         &0.0001             \\
-2	&0.0002 	&-		&L.L.	            \\
-3	&0.0097 	&0.0104		&L.L.		    \\
-4	&-		&L.L.		&-		    \\
\end{tabular}
\end{ruledtabular}
\end{table}

\subsection{Charge density differences.}

In figure \ref{fig Q=+1-1} we show the instantaneous change in charge 
density when an electron is added to (upper row) or removed from 
(lower row) the $Q=0$ charge state.  In other words, we first take the 
$Q=0$ relaxed positions and calculate with the $Q=-1$ or $Q=+1$ total 
charge.  We then subtract from this new charge density the original 
charge density from the relaxed $Q=0$ calculation.  The isocharge 
surfaces plotted are for 80\%, 50\% and 20\% of the peak value.  It is 
clear that an electron added goes to a localised orbital on or near 
the In triangle.  An electron removed, however, comes from a 
delocalised state.  We note, however, that even for the 20\% peak 
value isocharge surface for 0/-1, the total charge enclosed corresponds to 
only 0.48 electrons.  Hence, even for the $Q=-1$ charge state, most of 
the added electron has an apparently delocalised character.  This is, 
however, a result of the use of a relatively small supercell: the 
In-In distance is 3.7 \AA ~\emph{within} the cell, but about 9.2 \AA ~ 
between cells.  This produces a significant fake dispersion for the 
defect levels, and means that the transfer levels given in table 
\ref{Levels} should not be trusted too heavily.

In figure \ref{fig Q=+1+0-1} we show the 50\% peak isocharge surface for 
the instantaneous charge density change upon adding or removing 
electrons from the symmetric structures for $Q=+1$, 0 and -1.  The 
orbital filled when adding an electron to $Q=0$ and that emptied when 
removing an electron from $Q=-1$ are essentially the same.  Clearly 
$Q=-1$ is a localised state of the \zvps complex.  Adding an electron 
to the $Q=-1$ state gives a very similar shaped charge density - 
the opposite spin state of the same orbital.

Removing an electron from $Q=0$ empties a delocalised valence band 
state, as can be anticipated from the transfer level being found inside 
the valence band in the previous section.  However, adding an electron 
to the (fully relaxed) $Q=+1$ fills an In triangle localised orbital, 
none the less.  Clearly $Q=+1$ also corresponds to a localised state 
of the complex.

Going beyond +1 and -1, figure \ref{fig Q=+1+0-1} also indicates a 
pattern which we in fact find for all charge states from +6 to -6: the 
electrons added or removed at the transfer levels +2/+1 and +1/0 come 
from essentially the same orbital, which we shall call $\Phi_{0}$.  
The same is also try for the orbitals corresponding to the transfer 
level pairs 0/-1 and -1/-2, ($\Phi_{1}$), -2/-3 and -3/-4 ($\Phi_{2}$) 
and -4/-5 with -5/-6 ($\Phi_{3}$) Rather than plot 
all of them we show only one example of each orbital (transfer level 
pair).  For the first three of these orbitals the shape is best seen 
in a top view, looking down the (111) axis.  This is shown in figure 
\ref{fig Q=+2 to -4} for the symmetric, dimerised and antidimerised 
structures.  A clear relationship emerges: the dimer orbital for 0/-1 
is very similar to the antidimer orbital for -2/-3.  Likewise, the 
dimer orbital for -2/-3 and the antidimer orbital for 0/-1 are the 
same.  Furthermore, a symmetric combination (addition) of these two 
would give an orbital very much like that of the symmetric -2/-3 
orbital.  An antisymmetric combination (subtraction) would give 
something similar to the symmetric orbital for 0/-1, although in this 
case the similarity is not complete.  This may be due to differences 
in the degrees of dimerisation or perhaps to limitations in the 
calculation.  Despite this, a clear relationship between the orbitals 
filled/emptied at these transfer levels in the three nearly degenerate 
structures is present and will be considered in more detail in section 
\ref{model}.

For the orbital associated with -4/-5 and -5/-6 we show a 
side view in figure \ref{fig Q=+3 and -5}.  It is the Zn $2p^z$ 
orbital, as expected based upon the structures reported in the 
previous section.  Confirming this now allows us to fully explain the 
volume dependence of the complex, as shown in table \ref{Q str table}.  
The volume is reduced monatonically from $Q=+2$ to $Q=-6$.  For 
$Q=-4\rightarrow -6$ this is due simply to the change from DX to 
non-DX - as the Zn $2p^z$ orbital fills up, the $sp^2$ bonding to the 
P atoms is reorganised, with a normal $sp^3$ dangling bond developing.  
This can then bond on a more or less equal footing with the In 
dangling bonds.  Then from $Q=+2\rightarrow -4$ the In-In distances 
are reduced as electrons enter the In dangling bonds, increasing the 
bonding between them.  The volume reduction, however, is due only 
partly to this: partly it is due to the concurrent reduction in Zn-In 
distances, even though no bonds form between the Zn and In until 
$Q=-5$.  This reduction is electrostatic in origin.  The empty Zn 
$2p^z$ orbital, together with the relatively high ionicity of the Zn-P 
bonds, (see figure \ref{fig Q=0_elf},) leaves the positive Zn core slightly less 
well screened along the direction pointing towards the three In atoms.  
The negative charge on the In atoms, meanwhile, increases as electrons 
are added, so the Zn-In distance is reduced electrostatically.  Thus 
the reduction in effective vacancy volume at the 0/-1 transfer level, 
believed to have been observed in the positron annihilation 
studies,\cite{Slotte} is due to a combination of the In-In bonding and 
Zn-In electrostatic attraction, rather than to a DX to non-DX 
transition.

Regarding the other structures in tables \ref{Q str table} and 
\ref{Metastr table}, we find that, as anticipated from the structures 
themselves, the orbitals filled by adding further electrons to \zvpmf 
or removing further electrons from \zvppb are all delocalised.  We 
show only one example in figure \ref{fig Q=+3 and -5}, namely removing 
an electron from +5.  The orbital filled by removing an electron to 
the metastable shrunken structure of $Q=+2$ (also in figure \ref{fig 
Q=+3 and -5}) is identical to that associated with removing an 
electron from the ground structure of $Q=+0$.  This confirms the 
explaination given above of these shrunken structures in terms of 
holes excited off into delocalised valence band states.

\section{\label{model}Structural model for the \zvps complex.}

\subsection{\label{symmetry}Symmetry considerations.}

Since, for $Q>-5$, the Zn is in a DX-like position and the Zn $2p^z$ is 
much higher in energy than the other orbitals involved, we 
omit it, and base our model on the assumption that all additional 
electrons added/removed from the complex go to/come from the triangle 
of $sp^3$ hybrid dangling bonds on the In atoms.  These start out life 
transforming as the $0^+$ (s) and $1^-$ (p) irreducible 
representations of the $O_3$ group for the free atoms, which descend 
to the (1 dimensional) $A_1$ and (3 dimensional) $T_{2}$ 
representations of the tetragonal group $T_d$ when placed inside a 
zinc blend structured crystal.  (See the level splitting diagram in 
figure \ref{fig splitting}.) The point group of a free vacancy 
(without the Zn) is the same - $T_d$ - and so the bound states inside 
the vacancy are known to also be described\cite{Vacancy_JT} by the 
same $A_1$ and $T_{2}$ representations of $T_d$ - giving a deep symmetric 
level and a three-fold degenerate excited level.  (Jahn-Teller 
distortions lifting this degeneracy lead to the negative U effects for 
the free \vp.) In \zvps the Zn further lowers the point group to 
$C_{3v}$.  The $A_1$ representation of $T_d$ is projected onto the 
$A_1$ representation of $C_{3v}$, whilst the $T_{2}$ representation 
splits, giving a second $A_1$ representation and a 2 dimensional $E$ 
representation.  The two $A_1$ representations can mix, so that the 
hybridisation on the Zn is reorganised, to give the high (unoccupied) 
Zn $2p^z$ orbital observed in our calculations, plus a new totally 
symmetric level constructed primarily from the three remaining In 
dangling bonds.  The $E$ representation corresponds to a two-fold 
degenerate excited state, also constructed primarily from In dangling 
bonds.  In the neutral state the lowest $A_{1}$ level is full and the 
others are empty.  For charge states $Q=-1\rightarrow -3$ this 
degeneracy must also be lifted, usually by a further structural 
Jahn-Teller effect.  This can happen in two ways - via the formation 
of one stronger or one weaker bond, leading to the $C_{1v}$ point 
group, or via a process reducing the symmetry to $C_3$, splitting the 
$E$ level into $l=\pm1$ components.  The latter requires a rigid 
rotation of the In triangle around the (111) axis, alternately 
shortening and lengthening the In-P bonds in the plane perpendicular 
to (111).  This is not likely to be favoured, however, since the 
energy gained would be very small.  Indeed, we do not see any tendency 
towards this in our ab initio structures.

For certain $Q$ it is also possible for the degeneracies to be lifted 
by spin interactions.  In the $Q=-1$ state, where the In triangle 
contains 3 electrons, it would be possible to localise one electron on 
each In - making the lower $A_1$ and $E$ represenations degenerate 
- and then form an $S=\frac{3}{2}$ high spin state.  This would 
require\cite{SiC_theory} the In $sp^3$ dangling bonds to be spatially 
rather small (localised) and hence local Coulomb interactions to 
be rather strong.  This does not appear to be the case, however.  In 
additional calculations using the Local Spin Density Approximation 
(LSDA) to search for it we find no evidence of such a state being 
stable. Similarly, in $Q=-2$, with 4 electrons present, an $S=1$ 
high spin state could be formed in the half filled $T_2$.  Again, we 
find no indications that this is stable.

Instead, it seems that the symmetric structures for $Q=-1\rightarrow 
-3$ are stablised by weak Coulomb interations with no localisation.  
We can estimate the differences between the classical spatial overlaps 
of the three orbitals.  Specifically, we take the charge contained in 
the overlap regions of the isocharge surfaces for charge density 
differences, (as plotted in figures \ref{fig Q=+1-1}-\ref{fig Q=+3 and 
-5})) renormalised to a whole electron to allow comparisons, and 
evaluate what fraction of the charge overlaps classically.  Using the 
50\% isocharge surfaces in figure \ref{fig Q=+2 to -4}, we find that 
the highest orbital (corresponding to transfer levels -2/-3 and -3/-4) 
has 1.58 times as much overlap with the lowest (transfer levels 
+2/+1 and +1/0) as the middle orbital (0/-1 and -1/-2) has with the 
lowest.  (The overlap between the middle and the highest is about 29
times larger again.) Using the 20\% isocharge surface, where more of 
the actual charge is accounted for, but where the spurious defect 
level dispersion is starting to wash out the effect, we find that the 
lowest orbital has 1.62 times as much overlap with the highest as it 
has with the middle one.  These numbers indicates that in 
the symmetric geometries the $E$ level is split into orbitals having 
less and more overlap with the deep $A_1$ level.  These levels are 
thus split by Coulomb repulsion from the electrons filling the deep 
$A_1$ level.

\subsection{A tight binding model.}

The Jahn-Teller behaviour found in the ab initio studies can be 
explained using a simple tight binding model.  To do this we need 
only to consider the leading order effects of changes in the In-In 
distances.  We make no assumptions yet about the form of the degenerate 
associated orbitals $\phi_i$ and we consider only the ``hopping'' 
overlap matrix elements $t_{ij}$ which move electrons between the 
orbitals on In$_{i}$ and on In$_{j}$.  The Hamiltonian is simply
\begin{equation}
{\hat H} = \sum_{i,j;\sigma} -t_{i,j} c^\dagger_{j,\sigma} c_{i,\sigma}
\label{H}\end{equation}
where $c^{\dagger}_{i,\sigma}$ ($c_{i,\sigma}$) creates(annihilates) a spin 
$\sigma$ electron at $i$, the sums being over the three 
In atoms.  We make the selection $t_{23} = t + \delta$ and $t_{12} = t 
- \frac{\delta}{2} = t_{13}$.  Hence $\delta > 0$ corresponds to 
forming a single In$_{2}$-In$_{3}$ dimer (plus two antidimers) and 
$\delta < 0 $ to a single In$_{2}$-In$_{3}$ antidimer (plus two 
dimers).  (The $-\frac{\delta}{2}$ terms balancing the +$\delta$ 
allow direct comparisons between $\delta > 0$, = 0 and $ < 0$.) We 
omit Coulomb repulsion (we assume that $U\ll t$) and also the 
spin-spin interactions which would lead to high spin states.  On the 
other hand, we will treat the symmetric structures as (potentially) 
metastable, thus implicitly assuming the presense of (at least a) 
vanishing Coulomb repulsion between the $E$ levels and the deep 
$A_{1}$ level.  More importantly, we also omit elastic effects from 
the lattice.  To leading order, $\delta$ is linear in the change in 
In-In distance, but the elastic energy is quadratic.  We can therefore 
write the elastic energy for an individual bond between In$_i$ and 
In$_j$ as $\lambda\delta_{ij}^2t$, where $\lambda$ is (related to) an 
elastic constant.  Hence for both the dimer and antidimer states 
described above the elastic energy is 
$\approx\lambda\delta^2t+2\lambda\left(\frac{\delta}{2}\right)^2t = 
\frac{3\lambda\delta^2t}{2}$.  However, we omit this term at present, 
solving only equation \ref{H}.

In the dimer case ($\delta > 0$) the solution is:
\begin{eqnarray}        E_0  =& -\frac{1}{2} - \frac{\delta}{2} - \frac{F(\delta)}{2} &\approx -2-\frac{\delta^2}{6}\nonumber\\
               		E_1  =& -\frac{1}{2} - \frac{\delta}{2} + \frac{F(\delta)}{2} &\approx 1-\delta+\frac{\delta^2}{6}\\
                        E_2  =& 1 + \delta&\qquad \nonumber
\label{Emodel}
\end{eqnarray}
in units of $t$, (expansions to $O(\delta^2)$ only) where
\begin{eqnarray} 
    F(\delta) &= &\sqrt{3}\sqrt{3-2\delta+\delta^2} \nonumber\\
              &\approx &3-\delta+\frac{\delta^2}{3} 
\end{eqnarray}
We write the wavefunctions in the form $\Phi = (a_1,a_2,a_3)$ where 
$\{a_i\}$ are the coefficients of $\{\phi_i\}$ in $\Phi$.  The 
wavefunctions associated with $E_0$, $E_1$ and $E_2$ are thus
\begin{eqnarray} 	
    \Phi_0  =& \left(\frac{1+\delta-F(\delta)}{\delta-2},1,1\right) 											
               &\approx\left(1-\frac{\delta}{2}-\frac{\delta^2}{12},1,1\right)\nonumber\\
	       \Phi_1  =& \left(\frac{1+\delta+F(\delta)}{\delta-2},1,1\right) 
		&\approx\left(-2-\delta-\frac{2\delta^2}{3},1,1\right)\\
\Phi_2 =& \left(0,1,-1\right)&\quad\nonumber
\end{eqnarray}
respectively. (Normalisation is omitted for clarity.) Hence, the ground 
state $E_0$ is totally bonding, (an $A_1$ irreducible representation 
of $C_{1v}$,) but has more weight upon the dimerised In$_{2}$-In$_{3}$ 
bond, thus corresponding to a dimer bonded to the single site 
In$_{1}$.  This is a very different result to that 
obtained\cite{Nieminen1,Nieminen2,OurVP} for the free \vp, where the 
lowest lying state was totally symmetric with equal weight on all 
sites.  This new result for \zvps indicates that a dimer (or 
antidimer) may form even with only one electron present and no 
degeneracy.  (The Jahn-Teller theorum does {\it not} state that 
symmetry {\it cannot} be reduced in the absence of degeneracy.) This 
possible symmetry lowering is weak, however: the energy gained is only 
$\sim\delta^2$, the same order as the elastic energy lost.  It comes 
about because the dimerisation/antidimerisation breaks the $E$ 
irreducible representation of $C_{3v}$ up into a $B_{1}$ and a further 
$A_{1}$ represention of $C_{1v}$.  This new $A_{1}$ representation can 
mix with the other ones, leading to a weak gain in energy, even though 
Jahn-Teller does not actually require it here.  Taking elastic energy 
into account, the total energy of the dimer structure for $Q=+1$ (with 
one electron) is
\begin{equation}
    E^{Q=+1} = -\frac{t}{2}\left(1 + \delta + F(\delta) - 
    3\lambda\delta^2\right)
\end{equation}
Taking the derivative of $E^{Q=+1}$ with respect to $\delta$ we find 
that the lowest energy solution has $\delta>0$ (is dimerised) as long 
as $0<\lambda<0.17$.  This appears to correspond to a weak value of 
the elastic constant.  For $Q=-1\rightarrow-3$ a stable dimer state exists for any value of 
$\lambda$ and has a lower energy than the symmetric structure.  Our 
DFT calcualtions indicate that the case of \zvps in InP is near to 
this critical value of $\lambda=0.17$: if we do the relaxation using a 
4x4x4 k-point grid we do not find a stable dimer or a stable 
antidimer.  However, if we use the less accurate 2x2x2 k-point grid we 
do indeed find a non-symmetric structural groundstate for $Q=+1$, 
namely a 2.1\% dimerised structure, with 
the symmetric structure an excitation 0.002 eV above.  For $Q=+0$ we 
never find stable dimerised or antidimerised structures.

Turning to the model excitations for the dimer structure, $E_{1}$ also 
corresponds to an In$_{2}$-In$_{3}$ dimer, (another $A_1$ 
representation,) but now anti-bonded to the single site, upon which 
most of the weight is placed.  $E_{2}$ is an antibonded dimer (an 
antidimer - a $B_{1}$ irreducible representation of $C_{1v}$) with no 
weight at all on the single site.  For the antidimer structure 
($\delta < 0$) the groundstate remains the same (at least for small 
$\delta$) but there is less weight on the dimer bond and more on the 
single site.  Also, the order of the excitations is reversed, with the 
antidimer state ($E_3$ above) becoming the lower of the two.  If we 
now assume a form for $\phi_i$ related to $sp^3$ hybrids we can see 
what sort of charge densities we expect to see for the $\Phi_{j}$, as 
sketched in figure \ref{fig model wavefunctions}.  

In the case of the 
excitations for the symmetric structure, we find
\begin{equation}
\Phi_{1,2} = \Phi_{\pm} = \left(1,e^{\pm\frac{2\pi}{3}i},e^{\pm\frac{4\pi}{3}i}\right)
\end{equation}
with the two being, of course, degenerate.  Unfortunately, describing the 
lifting of this degeneracy as a result of Coulomb and elastic effects 
(the two will almost certainly combine) lies well outside the scope of 
this model or paper.  Hence what is shown in the figure is just the 
two alternatives with $C_3$ symmetry and three node-like planes, based 
upon the form of $\Phi_{\pm}$.  The one with low charge density along 
the In-In lines is likely to have a smaller overlap with - and hence 
Coulomb repulsion from - the symmetric $E_{0}$ groundstate.  Hence it 
has been drawn as the lower of the two excitations.  This cannot be 
predicted with more certainty upon the basis of the current model 
calculation, however.

With the obvious exception of the excitations for the symmetric 
structure, the similarity between the model charge densities in figure 
\ref{fig model wavefunctions} and the ab initio charge densities in 
figure \ref{fig Q=+2 to -4} is obvious.  We are indeed seeing 
the predicted combinations of one electron orbitals from the 
three dangling In dangling bonds, controlled by Jahn-Teller 
effects which the model can also predict.

Using equation \ref{Emodel} we can now evaluate the energy of the 
symmetric, dimerised and antidimerised states for each value of 
electron filling and compare them.  For example, with five electrons 
present (\zvpmc) the symmetric state has energy $-t$, the dimerised 
state has energy $-t(1+\delta)$ and the antidimerised state has energy 
$-\frac{t}{2}(1+\delta+F(\delta))\approx(-t(1+\delta+\frac{\delta^2}{6})$. 
 Hence the lowest is the antidimer, with the dimer as an excitation 
only $\sim \frac{\delta^2t}{6}$ higher, and the symmetric state lies 
$\sim \delta t$ higher still.  Comparing to table \ref{Excite ab 
initio table} we indeed see that the antidimer is the most stable, 
followed by the dimer, with the symmetric structure the least stable.  
Table \ref{Excite model} shows the leading terms in these excitation 
energies for fillings of 0 to 6 electrons.  Since both elastic and 
Coulomb energies favour the symmetric structure we anticipate only the 
symmetric structure for $Q=+2$ and $-4$, as observed.  We also 
anticipate that the excitation energies for the symmetric forms will 
be lower than those indicated in the table.  We find that for 1 or 2 
electrons the dimerised structure to be the most stable, 
closely followed by the antidimerised structure.  For 3, 4 or 5 
electrons we expect the antidimerised structure to be the lowest, with 
the symmetric structure always the least.  The agreement with the ab 
initio results is very close - the only real exception is that for 
$Q=-3$ our ab initio calculations found the dimerised state to be 
slightly more stable than the antidimerised.  The ab initio energy 
difference was tiny, however, over an order of magnitude less than 
the accuracy anticipated from such a calculation.  On the other hand, 
the model does predict that the symmetric structure is at its least 
stable for charge $Q=-2$ and this is indeed the only charge state for 
which we are unable to find the symmetric structure using DFT.

\begin{table}
\vglue -0.3truecm
\caption{\label{Excite model}Model predictions for the lowest lying 
structures (marked L.L.) of \zvpq, together with structural excitation 
energies to leading order in $\delta$, for occupancies of 0 to 6 
electrons (\# $e^{-}$).  DM = dimerised, SY = symmetric, AD = 
antidimerised.}
\begin{ruledtabular}
\begin{tabular}{crccc}
\# $e^{-}$      &Q      &DM                     &SY                     &AD \\\hline
0		&+2	&-		        &L.L.		        &-			\\
1		&+1	&L.L.			&$\frac{\delta^2t}{6}$	&$\frac{\delta^3t}{9}$	\\
2		&0	&L.L.			&$\frac{\delta^2t}{3}$	&$\frac{2\delta^3t}{9}$	\\
3		&-1	&$\frac{\delta^2t}{6}$	&$\delta t$		&L.L.			\\
4		&-2	&$\frac{\delta^2t}{3}$	&$2\delta t$		&L.L.			\\
5		&-3	&$\frac{\delta^2t}{6}$	&$\delta t$		&L.L.			\\
6		&-4	&-			&L.L.			&-			\\
\end{tabular}
\end{ruledtabular}
\end{table}

\section{\label{conclusions}Conclusions.}

We have calculated the structure and transfer levels of the \zvps 
complex using ab initio DFT methods and have presented detailed 
results for the dependence of the structure and structural excitations 
upon the charge state of the complex and hence upon the Fermi level.
We found that, for all charge states below -5, the Zn atom lies in a 
DX position, $sp^2$ hybridised, with an empty $2p^z$ orbital 
directed towards the In triangle and very constant bond lengths to the 
surrounding P atoms.  Electrons added (below $Q=-5$) go into localised 
orbtials built from the three In dangling bonds - thus reducing the 
In-In distances with each added electron.  We also found that there is 
a simultaneous reduction in the Zn-In distances, which is of 
electrostatic in origin.  This was due to the increased charge on the 
In triangle and the weaker screening of the Zn core along (111) owing 
to the empty Zn $2p^z$ orbital.  The combination of these effects 
leads to a systematic reduction in the effective vacancy volume of the 
complex for charge states from $Q=+2\rightarrow-4$.  This is in spite 
of the fact that the DX structure of the complex remains stable with 
rising Fermi level right up until a -4/-6 negative $U$ transfer level, 
about 0.5 eV into the conduction band.  For charge states 
$Q=-1\rightarrow-3$ we also found complex Jahn-Teller behaviour, with 
up to three different stable structures - symmetric, dimerised and 
antidimerised - which are all very close to degeneracy.  For $Q=-1$ 
the most stable structure (just) is dimerised, whilst it is 
antidimerised for $Q=-2$ and -3.

In addition, we have presented a simple tight binding model which 
is able to predict and explain all of the structural properties of the 
complex - including the existance of three possible nearly degenerate 
structures for $Q=-1\rightarrow-3$.  Such a near degeneracy between 
dimer and antidimer structures has not been seen for the free \vps
and this too the model can predict: the same model applied to \vpma, 
for example, shows a separation between dimerised and antidimerised 
structures on the order of $\delta$, compared to $\delta^2$ here for 
\zvp.  (Details of these calculations will be presented 
elsewhere.\cite{OurVP}) Success with such a simple model is very 
pleasing and may in the future even help to narrow the range of initial 
configurations required when searching for structural minima in 
similar ab initio calculations.

Comparing our results to those obtained experimentally using positron 
annihilation, we, as they, find that the most stable charge state of 
the \zvps complex is $Q=0$ in p-type material.  We find a binding energy 
of 0.39 eV which is also in the range they anticipate.  Regarding the 
structure of the complex, we, as they, find a larger effective vacancy 
volume than that of the free \vp.  Upon excitation to the -1 charge 
state we, as they, see a reduction in volume.  At 0.50 eV our 0/-1 
transfer level is a little high for agreement, but, as was noted, 
these transfer levels are not reliable since the supercell is too 
small.  Further studies of the complex in larger cells are required to 
check how this effects the position of the transfer levels, but our 
results seem otherwise to be in very good agreement with the 
experiments.

Regarding the issue of compensation, our results confirm that, should 
free \vps somehow be present in heavily Zn-doped InP, they would 
indeed form stable complexes with Zn.  However, the formation energy we 
find for \vps is 1.96 eV, which is still rather large.  Even for a 
phosphorus atom neighbouring a Zn dopant the effective formation enery 
is 1.96 - 0.39 = 1.57 eV, which is not small.  The positron 
annihilation experiments clearly demonstrate that at least some \vps 
are present in heavily Zn doped Czochralski grown InP.  However, the 
lack of a low formation energy leaves it unclear whether, and perhaps 
unlikely that, \vps generation is the primary cause of 
compensation in InP:Zn.  On the other hand, our DFT results indicate 
that the formation of \zvps complexes from existing \zins and \vps is 
certainly {\it not} responsible for compensation: starting with fully 
ionised Zn, the complex forming reaction is
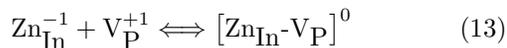
\begin{equation}
\hbox{Zn}_{\hbox{In}}^{-1} + \hbox{V}_{\hbox{P}}^{+1} 
\Longleftrightarrow 
\left[\hbox{Zn}_{\hbox{In}}\hbox{-}\hbox{V}_{\hbox{P}}\right]^0
\end{equation}
in which no free electrons or holes are created or absorbed.  The 
compensation issue here has to do with the formation of the phosphorus 
vacancies in the first place, and is not affected by the formation of 
the \zvps complex.  Fully understanding the issue of carrier 
compensation in InP:Zn from an ab initio point of view may thus 
require more work.

\section{Figure Captions.}

\subsection{Figure 1.}
Structure of the supercell containing the 
neutral \zvps complex.  Atom types are dark grey: In, pale grey: P, 
black: Zn.

\subsection{Figure 2.}
ELF plots for the neutral \zvps 
complex.  For clarity most atoms are omitted and the edges of the 
Zn-In tetrahedron are drawn.  Note, however, that there are \emph{no} Zn-In bonds.  a) 
ELF in a plane containing Zn, 1 In and the (111) axis. b) ELF in a plane 
containing Zn and 2 In.

\subsection{Figure 3.}
Charge density difference plots for the 
\zvps complex in the $Q=0$ relaxed geometry, but with an extra 
electron added (upper row) or removed (lower row).  Shown are 
isocharge surfaces for the charge density difference itself.  The 
surfaces shown are for 80\%, 50\% and 20\% of the peak 
value.

\subsection{Figure 4.}
50\% isosurface plots for charge 
density differences for adding (top row) and removing (bottom row) 
electrons from the$Q=+1$, 0 and -1 relaxed structures for \zvp.

\subsection{Figure 5.}
Top views (down the (111) axis) of the 
50\% isosurface plots for charge density differences corresponding to 
the transfer levels of \zvps from $Q=+2$ to -4, for the symmetric, 
dimerised and antidimerised structures.

\subsection{Figure 6.}
Side 
views of charge density difference plots for a) the Zn $2p^z$ related 
orbital filled at the -4/-5 and -5/-6 transfer levels, b) the 
delocalised orbital emptied in removing an electron from $Q=+5$ and c) 
the orbital emptied in removing an electron from the metastable 
shrunken structure for $Q=+2$.

\subsection{Figure 7.}
Irreducible representation splittings as symmetry is 
lifted from $O_3$ to $C_{1v}$ or $C_{3}$.  For $C_{1v}$ the dimer 
case is shown.  For the antidimer the $B_{1}$ level lies below the second 
$A_{1}$ level.

\subsection{Figure 8.}
Sketch of the wavefunctions anticipated for the three 
structures of \zvp, based upon the model. (DM = dimerised, SY = symmetric, AD = 
antidimerised.)

\begin{figure}
\caption{\label{fig Q=0_st}}
\end{figure}

\begin{figure}
\caption{\label{fig Q=0_elf}}
\end{figure}

\begin{figure*}
\caption{\label{fig Q=+1-1}}
\end{figure*}

\begin{figure*}
\caption{\label{fig Q=+1+0-1}}
\end{figure*}

\begin{figure*}
\caption{\label{fig Q=+2 to -4}}
\end{figure*}

\begin{figure*}
\caption{\label{fig Q=+3 and -5}}
\end{figure*}

\begin{figure}
\caption{\label{fig splitting}}
\end{figure}

\begin{figure}
\caption{\label{fig model wavefunctions}}
\end{figure}

\bibliography{eprint}
\end{document}